\documentclass[conference]{IEEEtran}
\IEEEoverridecommandlockouts

\usepackage{amsmath, amssymb}
\usepackage{bm}
\usepackage{algorithm}
\usepackage{algorithmic}
\usepackage{pifont}
\usepackage{makecell}
\usepackage{multirow}
\usepackage{array}
\usepackage{tabularx}
\usepackage{booktabs}
\usepackage{graphicx}
\usepackage{xcolor}
\usepackage{cite}
\usepackage{url}
\usepackage{hyperref}
\hypersetup{pdftitle={Network-in-the-Loop at Scale: GPU-Batched 5G Simulation for Massively Parallel Robot Learning}, pdfauthor={Zifan Zhang, Mingzhe Han, Kannan Athreya, Yuchen Liu}}
\usepackage{xspace}

\newcommand{\myparatight}[1]{\smallskip\noindent{\bf {#1}:}~}
\newcommand{\alg}{Isaac-Net\xspace}

\title{Network-in-the-Loop at Scale: GPU-Batched 5G Simulation for Massively Parallel Robot Learning}

\author{%
\IEEEauthorblockN{Zifan Zhang\IEEEauthorrefmark{1}\IEEEauthorrefmark{3}, Mingzhe Han\IEEEauthorrefmark{2}\IEEEauthorrefmark{3}, Kannan Athreya\IEEEauthorrefmark{2}, and Yuchen Liu\IEEEauthorrefmark{1}}
\IEEEauthorblockA{\IEEEauthorrefmark{1}Department of Computer Science, North Carolina State University, Raleigh, NC, USA\\
\IEEEauthorrefmark{2}Department of Computer Science and Engineering, The Ohio State University, Columbus, OH, USA\\
Email: \{zzhang66, yuchen.liu\}@ncsu.edu, \{han.1453, kannan\}@cse.ohio-state.edu}
\thanks{\IEEEauthorrefmark{3}These authors contributed equally to this work.}}

\begin{document}

\maketitle

\begin{abstract}
Massively parallel GPU simulators train multi-robot policies in thousands of environments, and many fleets use private Fifth-Generation (5G) networks, where each robot's delay depends on its teammates' traffic.
Network-in-the-loop training places a simulated 5G network inside this loop.
However, GPU robot simulators reduce the network to an independent delay per message, while packet-level simulators run one scenario per CPU process and cannot keep pace with thousands of parallel environments.
To bridge this gap, we present \alg, a GPU-batched 5G New Radio (NR) module that advances the uplink of thousands of environments in lockstep with Isaac Lab physics.
\alg simulates every slot, the 0.5~ms interval in which the base station decides which robots transmit, for all environments at once.
Extensive experiments confirm that its NR engine reproduces the median delay of ns-3 5G-LENA across loads, with a median delay 5--10\% low on an unseen carrier and 9\% high at 32 robots per environment in closed loop.
The engine also reproduces the Age of Information (AoI), the age of each robot's newest delivered report, while an independent delay per message leaves the AoI tail about three times too light.
In a configuration validated against 5G-LENA, \alg keeps the network in the loop for about one million robots on one GPU at 83\% of the Isaac Lab rate without the network, measured under a random policy.
\alg is open source at \url{https://github.com/ZzZTripleZzZ/isaac-net}.
\end{abstract}

\section{Introduction}
\label{sec:intro}

Robot learning has moved onto the GPU, where simulators such as Isaac Gym~\cite{makoviychuk2021isaacgym}, Isaac Lab~\cite{mittal2025isaaclab}, ManiSkill3~\cite{tao2024maniskill3}, and MuJoCo Playground~\cite{zakka2025playground} step thousands of environments in parallel and feed the resulting experience directly into reinforcement learning.
This throughput has made it practical to train policies for whole teams of robots, such as warehouse fleets, drone swarms, and multi-arm workstations, in which each environment holds tens to hundreds of agents~\cite{bettini2022vmas,lan2021warpdrive}.
Many of these teams report sensor data to an edge server and receive commands from it over a wireless network.
Moreover, private Fifth-Generation (5G) networks are increasingly deployed for mobile robots in factories and campuses~\cite{sanchez2025ndt5grobots,raghavendra2026isaacsim5g}.
A policy trained for such a team therefore acts on information that the network may have delayed or lost, so the network belongs inside the massively parallel training loop.
We call such training network-in-the-loop, by analogy with hardware-in-the-loop testing, although the network inside the loop is simulated, not a real one.

In a 5G cell, the base station, called the gNB, divides one pool of radio resources among the robots in every slot.
A slot is a fixed scheduling interval, 0.5~ms at the common 30~kHz subcarrier spacing, in which the gNB decides which robots transmit based on channel quality and on Buffer Status Reports (BSRs), the reports in which robots state how many bytes they hold.
A robot may transmit only after the gNB sends it a grant, that is, a permission to transmit on given resources, which the gNB issues after a scheduling request and then sizes from later buffer reports.
A failed transmission is retried several slots later through Hybrid Automatic Repeat reQuest (HARQ), a fast retransmission scheme of the radio link, and unsent messages wait in per-robot queues.
Consequently, the delay that a robot experiences depends on its own position and on what its teammates in the cell transmit, meaning that network behavior and team behavior are coupled.
This coupling also closes a loop between the policy and the network.
For instance, a policy may send more reports when reports arrive late, which could congest its own uplink and delay every later report.
A learning environment that captures this loop could teach a fleet when to send, how much to send, and where to move so that critical reports arrive in time.
However, GPU-parallel robot simulators typically model the network, if at all, as a fixed delay or as a delay drawn independently for each message.
For instance, the delayed actuator model of Isaac Lab~\cite{mittal2025isaaclab} applies each command a set number of physics steps late, so it cannot express contention among teammates.

\begin{figure}[t]
\centering
\includegraphics[width=\columnwidth]{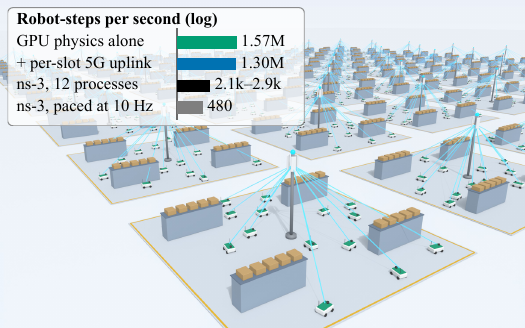}
\caption{Parallel warehouse environments whose robot fleets each send uplink traffic to a simulated gNB that schedules every slot, with an inset of robot-steps per second for GPU physics alone, with the 5G uplink, and for two ns-3 5G-LENA setups (log scale, higher is better).}
\label{fig:teaser}
\end{figure}

\myparatight{Motivation and Challenges}
Putting a faithful 5G network inside a GPU-parallel training loop raises three challenges.
We measure simulation speed in robot-steps per second, where a robot-step advances one robot by one control step, the 100~ms interval at which the policy acts.
\textit{First}, delays from contention among robots for the same radio resources, from HARQ retransmissions, and from scheduling arise per slot and persist across a robot's successive messages.
Thus, an independent delay per message misses part of what a controller observes under load.
\textit{Second}, a packet-level simulator on a CPU advances one scenario at a time, whereas GPU physics steps thousands of environments at once, so the network would stall the training loop.
Figure~\ref{fig:teaser} shows that a pool of 12 ns-3 5G-LENA~\cite{patriciello2019lena} processes, which run the 5G module of the packet-level simulator ns-3, delivers a few thousand robot-steps per second on a shared CPU.
One ns-3 process that advances in lockstep with the controller at 10~Hz keeps pace for only 48 robots.
In contrast, Isaac Lab physics alone reaches about 1.6 million robot-steps per second with simple sphere-shaped robots on one GPU.
\textit{Lastly}, the network must stay aligned with 20~ms physics substeps and with control steps while Isaac Lab resets any subset of environments at any step.
A simulator that pauses or restarts as a whole would therefore hold back every other environment.

\myparatight{Limitations and Opportunities}
Mature mobile network simulators, such as ns-3 with its LTE and 5G-LENA~\cite{patriciello2019lena} modules, Simu5G~\cite{nardini2020simu5g}, the Vienna 5G simulators~\cite{muller2018vienna5g}, and the MATLAB 5G Toolbox, model cellular networks accurately, and Veins~\cite{sommer2011veins} couples such simulators with road traffic.
Recent works have also coupled robot simulators with network simulators~\cite{calvofullana2021rosnetsim,acharya2022cornet2,dey2022synchrosim,cauwe2025gazebons3,yan2026simart,raghavendra2026isaacsim5g}, yet each typically pairs one robot scene with one network instance through the Robot Operating System (ROS) or a real-time broker, so it runs at or below wall-clock speed.
GPU network simulators such as GeDES~\cite{li2026gedes} accelerate one large wired network.
Sionna SYS~\cite{sionna2026sys} batches the 5G link model and scheduler, but it keeps no queue per robot, no HARQ timing, and no uplink access driven by buffer reports.
QNet~\cite{agarwal2026qnet} learns when an agent should send over a shared network, yet its training network has no per-slot scheduling or HARQ and is not coupled to robot physics.

GPU-parallel reinforcement learning created three requirements that none of these tools meets together: 1) a batch dimension, so that thousands of independent scenarios advance together on one GPU; 2) lockstep with a physics engine, in which the network advances exactly one control step whenever the physics does, returns per-robot observations as tensors, and resets single environments at arbitrary times; and 3) closed-loop traffic and mobility, in which the policy decides what each robot sends and the physics decides where it is, so nothing can be precomputed or replayed.
Training experiments also need every environment to stay reproducible, which calls for a separate random stream per environment so that its results do not change with the batch size.
The network model must also avoid host synchronization, that is, it must never make the GPU wait for the CPU.
These requirements appeared with GPU-parallel learning, and \alg is built around them from the ground up.

\myparatight{Contributions}
To this end, we develop \textbf{\alg}\footnote{Project page: \url{https://isaacnet.zifanzhang.com}. Code: \url{https://github.com/ZzZTripleZzZ/isaac-net}.}, a GPU-batched 5G New Radio (NR) module that brings network-in-the-loop training to Isaac Lab at full parallel scale, where the robots of every environment contend slot by slot for the uplink of a shared cell.
In essence, three core techniques are developed:
\begin{itemize}
\item \textit{First}, \alg stores each robot's channel, scheduler state, HARQ processes, access timers, and message queue in fixed-shape tensors indexed by environment and robot. The same tensor operations then advance the state of all robots in all environments at every slot. Per-environment clocks, partial resets, and random draws keyed by environment and step let Isaac Lab restart any subset of environments while every other environment continues bitwise unaffected.
\item \textit{Second}, a configurable NR engine (level L2 in the code) models 5G scheduling, decoding errors, retransmissions, and uplink grants, while a simpler prototype engine (level L2-legacy in the code) remains as a cheaper option. Both engines run on three backends, namely a plain PyTorch reference, a graph backend that is bitwise identical to it, and a triton backend that fuses one control step into a single GPU function, called a kernel, and matches the reference to floating-point rounding. Equivalence tests feed identical random draws to a fast backend and the reference and compare every output and state value at every step, thereby verifying each speed-up before it runs at the scale of millions of robots.
\item \textit{Third}, a mixin class, that is, a reusable add-on class, inserts the network through four hooks into any direct-workflow Isaac Lab environment, in which one class computes observations and rewards. It returns each robot's Age of Information (AoI)~\cite{kaul2012aoi}, that is, how old its newest delivered message is, together with its queue length, Signal-to-Noise Ratio (SNR), and delivery flag. An Isaac Lab event term, a function called at every reset, redraws the radio parameters, so the policy trains over many radio conditions, which is called domain randomization.
\item \alg is released as the pip-installable package \texttt{isaac-net}, with its test suite, co-simulation bridges to ns-3 5G-LENA and to the OpenAirInterface (OAI) 5G stack~\cite{nikaein2014oai}, an Isaac Lab fleet task, and documentation on the project page.
\item Extensive experiments show that, in the scale configuration, that is, the validated NR configuration without the buffer-report grant procedure, Isaac Lab runs 1,048,576 robots with the network in the loop at 1.24 control steps per second on one GPU under a random policy, against 1.50 without the network. With that procedure and no fitted parameter, the NR engine reproduces the per-run median delay of 5G-LENA within a median absolute error of 2.5--4.8\% from light to saturated load. Under load in closed loop, it also reproduces the AoI that a hand-written controller, not a learned policy, observes over ns-3, while an independent delay per message, even one drawn from the engine's own delays, leaves the AoI tail about three times too light.
\end{itemize}

To the best of our knowledge, \alg is the first 5G network model that runs batched on the GPU inside a massively parallel robot learning loop.

\section{Design of \alg}
\label{sec:design}

\subsection{Problem Setting}
\label{sec:setting}

\alg serves $E$ parallel environments with $R$ robots each, where every robot is a 5G User Equipment (UE) attached to the gNB of its environment.
All robots of one environment share the radio resources of this gNB.
Isaac Lab owns the physics, the policy, and the reward, while \alg owns the path of each message from the robot to the edge server.
Time advances in control steps of length $\Delta$, the interval at which the policy acts, and each control step covers $K$ uplink slots of the NR frame structure.
The goal is to advance all $E$ networks by one control step at a cost that is small compared with the physics of that step, while each environment evolves as if simulated alone slot by slot.

However, this problem is intrinsically difficult for three reasons.
First, the Medium Access Control (MAC) layer couples all robots of a cell at every slot, which rules out simulating robots independently.
Second, the clocks of radio slots, physics substeps, and control steps differ by two orders of magnitude, which rules out advancing the network only once per control step.
Lastly, Isaac Lab ends and restarts episodes of different environments at different times, which rules out restarting the whole simulation, as a packet-level simulator would.
These factors must be handled for thousands of environments within every physics step, which makes a direct GPU port of an event-driven simulator impractical and motivates the following design.

\subsection{Overview of \alg}
\label{sec:overview}

\begin{figure*}[t]
\centering
\includegraphics[width=\textwidth]{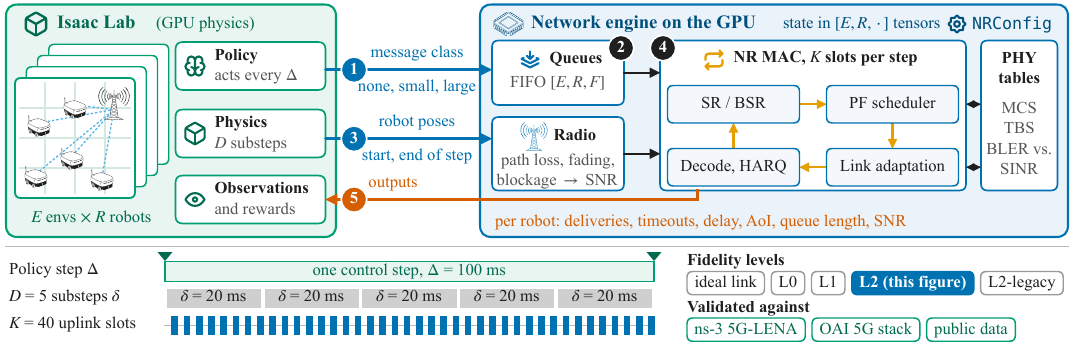}
\caption{Overview of \alg over one control step, from message classes and robot poses as inputs, through $K$ uplink slots of the NR MAC on GPU tensors, to per-robot deliveries, delays, AoI, queue lengths, and SNR as outputs.}
\label{fig:overview}
\end{figure*}

\alg advances the networks of all $E$ environments together, as one batch of tensor operations over the $E \times R$ robots, so a single call serves the whole parallel training loop.
It exposes two API calls on tensors, where \texttt{submit} takes the message class that each robot sends in this step, and \texttt{step} takes the robot poses and returns what the network delivered.
Two per-slot engines implement this interface, the configurable NR engine and a prototype engine, which the following Section~\ref{sec:levels} compares.
\textit{A High-Level Walkthrough:} Figure~\ref{fig:overview} traces one control step through \alg in five numbered steps.
In step~1, the policy submits a message class per robot, namely nothing, a small message such as a status report, or a large message such as a camera frame.
In step~2, the message queue captures each message with the robot's pose at the start of the step and appends it to that robot's First-In First-Out (FIFO) buffer.
In step~3, after the physics substeps, the radio model converts the start-of-step and end-of-step poses into a per-robot SNR.
In step~4, the NR MAC runs all $K$ uplink slots of the step for the robots of all $E$ environments at once.
In each slot, every gNB schedules its robots, selects a Modulation and Coding Scheme (MCS), and decodes Transport Blocks (TBs) against the Physical Layer (PHY) tables with HARQ retransmissions.
In step~5, \alg returns per robot the delivered and dropped messages, their delay, the AoI, the queue length, and the SNR, which the environment turns into observations and rewards.
The strip at the bottom of the figure shows how one control step of length $\Delta$ aligns $D$ physics substeps with $K$ uplink slots.
The boxes at the bottom right of the figure list the fidelity levels, that is, network models of increasing detail described in Section~\ref{sec:levels}, and the simulators that \alg was validated against.

\subsection{Time-Stepped Slot Loop}
\label{sec:slots}

Packet-level simulators such as ns-3 process events one at a time in time order, which batches poorly because events of different environments fall at different times and follow different code paths.
\alg instead exploits a property of the NR MAC layer, namely that scheduling, transmission, decoding, and HARQ feedback all take effect at slot boundaries.
A time-stepped loop over slots therefore visits every instant at which the modeled MAC state can change, and it needs no event queue, no dynamic memory, and no code path that depends on which environment is being processed.
Each slot becomes a fixed sequence of tensor operations over the leading dimensions $[E, R]$, which a GPU executes efficiently.
A single fused kernel can run this sequence, and so can a CUDA graph~\cite{nvidia2026cudaguide}, a recorded chain of kernels that the GPU replays with a single launch.
In the Time-Division Duplex (TDD) pattern of the frame, only some slots carry uplink traffic, so a control step of $\Delta$ (e.g., $100$~ms) spans $K$ (e.g., $40$) uplink slots.

Two approximations make the loop efficient.
First, messages enter the network at control-step boundaries, which matches a policy that decides once per step what to send.
Second, the radio averages the SNR over $J$ (e.g., $4$) poses interpolated across the step, while fast fading, the rapid fluctuation of the signal, still evolves every slot inside the engine.
Since mobile robots typically move at a few meters per second, the geometry changes by well under a meter within one step.

\subsection{Fixed-Shape Network State}
\label{sec:state}

Every state variable of \alg is a tensor with leading dimensions $[E, R]$ and a fixed trailing shape.
For instance, $[E, R, S]$ holds the fading of $S$ subbands, the frequency blocks of the carrier, $[E, R, H]$ the $H$ HARQ processes, and $[E, R, F]$ a FIFO of $F$ messages per robot.
Algorithm~\ref{alg:step} lists one control step on this state.

\begin{algorithm}[t]
\caption{One control step of the \alg NR engine.}
\label{alg:step}
\begin{algorithmic}[1]
\REQUIRE message classes $\bm{q}\in\{0,1,\dots\}^{E\times R}$, start and end poses $\bm{P}^{\mathrm{s}}, \bm{P}^{\mathrm{e}}\in\mathbb{R}^{E\times R\times 3}$
\STATE Write each new message into FIFO entry \texttt{count} with a one-hot mask, recording its size and capture step
\STATE $\bm{\gamma} \leftarrow \frac{1}{J}\sum_{j=1}^{J}\mathrm{SNR}\big(\mathrm{interp}(\bm{P}^{\mathrm{s}},\bm{P}^{\mathrm{e}}, j/J)\big)$
\FOR{$k = 1$ \TO $K$}
\STATE Update the fast fading of every robot and subband
\STATE Advance Scheduling Request (SR) timers, update BSRs, and admit grants whose delay has elapsed
\STATE Schedule pending retransmissions first, then the remaining resources by Proportional Fair (PF)~\cite{jalali2000pf} priority
\STATE Select each MCS from the effective Signal-to-Interference-plus-Noise Ratio (SINR) and the outer-loop link adaptation offset, and size each TB
\STATE Decode each TB with a Block Error Rate (BLER) draw, then update HARQ processes and the link adaptation offset
\STATE Serve the bytes of decoded TBs from the FIFO in order and stamp completed messages with slot $k$
\ENDFOR
\STATE Drop messages older than the timeout and compute delivered masks, delays, AoI, and queue lengths
\RETURN outputs and the updated state
\end{algorithmic}
\end{algorithm}

\textbf{Stage I (Enqueue): } The message queue writes new messages into the first free FIFO entry of each robot through a one-hot mask, a mask with a single 1 at the target entry (Line 1).
The mask avoids the GPU-to-CPU wait that an index lookup such as \texttt{nonzero} would force, and finished messages are removed while the remaining ones shift forward in order.
\textbf{Stage II (Radio): } The radio converts the interpolated poses into one SNR per robot and cell from path loss, correlated shadowing, and optional blockage by other robots (Line 2).
\textbf{Stage III (Slot loop): } The MAC layer advances fading, access, scheduling, link adaptation, decoding, and HARQ for all robots at once, and it serves queued bytes in order (Lines 3--10).
Because the FIFO holds bytes in a stream, a grant of $c$ bytes serves $s_i$ bytes of message $i$, which has $r_i$ bytes left and starts $b_i$ bytes into the grant, given by
\begin{equation}
\label{eq:fifo}
s_i = \max(c - b_i, 0) - \max(c - r_i - b_i, 0),
\end{equation}
which equals $r_i$ exactly once the grant covers the whole message, so a tiny rounding leftover can never keep a fully served message from completing.
\textbf{Stage IV (Outputs): } The queue drops messages that exceed the application timeout and derives the outputs that the environment observes (Lines 11--12).

Isaac Lab resets environments individually, so every state tensor supports a partial reset.
Thus, \alg keeps a per-environment clock, defined as the global slot clock minus the step at which the environment was last reset, and a reset rewrites only the rows of the selected environments in place.
Every random number is computed by hashing the seed, the environment, the episode, the step since reset, the slot, and the place in the code that draws it.
Thus, an environment's draws depend neither on $E$ nor on other resets, and a reset leaves every other environment bitwise unaffected and never reallocates a buffer.

\subsection{Backends and the Equivalence Guarantee}
\label{sec:backends}

Fixed shapes make an engine fast only if its per-slot operations are fused.
At $E \times R = 64 \times 16$, the prototype engine in plain PyTorch launches about 3,000 small kernels per control step and the NR engine about 27,000, each doing only microseconds of work.
\alg therefore provides three backends that share the same state and API for both engines.
The \textit{reference} backend implements the model in plain PyTorch for readability and thereby defines it.
The \textit{graph} backend keeps every state tensor in a persistent buffer, derives slot and HARQ times from a device-side step counter, and captures one control step as a CUDA graph that is replayed without Python overhead.
The \textit{triton} backend runs all slots of a control step in one Triton~\cite{tillet2019triton} kernel with one GPU program per environment, which keeps each robot's state in registers, the fastest on-chip memory, and evaluates the same hash as the reference.
We call the graph and triton backends the fast backends.
The triton kernel of the NR engine covers the whole slot, from fading to the Radio Link Control (RLC) layer, and runs the downlink as a second kernel.
It is single-cell and supports every option of the \texttt{lena\_validation\_v2} preset of Section~\ref{sec:nrconfig} except the buffer-report grant procedure, in which buffer reports size the later grants.
The scale experiments therefore use this preset with a single fixed access delay, the time from a scheduling request to the first uplink data, in place of that procedure.
This preset is the scale configuration, and Section~\ref{sec:eval-lena} validates it against 5G-LENA.

Such optimizations can silently change the model, so \alg treats every fast backend as a refactoring that must be verified against the reference.
The equivalence tests feed the same pre-generated decode draws and fading innovations, that is, random increments of the fading process, to both backends and compare every step output, per-message field, and MAC state tensor with exact equality.
The tests run in a full-run mode, in which each backend evolves its own state, and in a per-step mode, in which both backends start from the same recorded state at every step because the reference state is copied into the fast backend before each step.
The per-step mode thus separates per-step disagreements from the later divergence that one early disagreement causes.
The graph backend must be bitwise identical in both modes, while the triton backend, which sums values in a different order and rounds differently, must agree to floating-point rounding.
Rare flips of a discrete decision, such as a TB decode at its threshold, are acceptable if they leave aggregate statistics unchanged.
A \texttt{torch.compile} backend of the prototype engine failed this test and is excluded from training runs (see Section~\ref{sec:eval-equiv} for details).

\subsection{Fidelity Levels}
\label{sec:levels}

Every model in \alg implements the same API, so a task selects the network model by changing one argument.
The i.i.d. delay model (level L0 in the code) draws a delay and a loss independently per message, as current robot simulators typically do, with the delay either log-normal or resampled from a given delay sample.
The fluid model (level L1 in the code) approximates the MAC by letting active robots share the Physical Resource Blocks (PRBs) equally in every slot while their FIFO queues drain.
The two most detailed levels run the configurable NR engine of Section~\ref{sec:nrconfig} and the frozen prototype engine, which ignores the MAC and PHY fields of the configuration.
The prototype engine uses a logistic BLER curve, one HARQ process, a fixed noise power, and a minimal grant after two uplink slots, and the NR engine replaces each of these simplifications with a 3GPP-based model.
Both engines have all three backends, and Section~\ref{sec:eval-lena} compares both with 5G-LENA.

\subsection{Configurable NR}
\label{sec:nrconfig}

One configuration object, \texttt{NRConfig}, configures every module of the NR engine, and Table~\ref{tab:config} summarizes what it exposes.
The frame structure follows the 3GPP specifications, and the Resource Block Groups (RBGs), MCS tables, and exact TB sizes follow TS~38.214~\cite{3gpp38214}.
The engine combines the SINR of all RBGs into one effective SINR by Exponential Effective SINR Mapping (EESM)~\cite{brueninghaus2005eesm}, which weights weak RBGs more, since they dominate decoding errors.
It then looks up the BLER in tables exported from Sionna~\cite{sionna2026sys} or generated locally from a 5G-LENA checkout~\cite{patriciello2019lena,lagen2020phyabstraction}.
A TB that exhausts its HARQ transmissions is either retried by the RLC layer in Acknowledged Mode (AM) or lost in Unacknowledged Mode (UM).
Outer-Loop Link Adaptation (OLLA)~\cite{nakamura2002olla} shifts the SINR estimate after every acknowledgment to absorb estimation errors, and SRs and BSRs follow the access procedure of TS~38.321~\cite{3gpp38321}.
Four configuration options, off by default and with no fitted parameter, reproduce how a 5G-LENA gNB serves a loaded cell.
First, the per-RBG PF update recomputes a robot's average throughput after every RBG it wins, which lowers its priority for the next RBG, so several backlogged robots split each slot.
This update also leaves the averages of idle robots unchanged.
Second, Time-Division Multiple Access (TDMA) retransmissions give every uplink retransmission a whole slot.
Third, link adaptation picks each robot's MCS assuming it will receive as many RBGs as in its previous uplink transmission, as 5G-LENA does.
Lastly, the buffer-report grant procedure replaces the single fixed access delay with a small first grant, after which buffer reports, which arrive late and in coarse size steps, size the later grants.
It also models the wait of a message's last bytes for an RLC status timer.
The NR engine also implements a downlink with gNB queues, channel-quality reports, and HARQ feedback, which the fast backends reproduce but which has not yet been compared with another simulator.

\begin{table}[t]
\centering
\caption{Network features selectable through \texttt{NRConfig}.}
\label{tab:config}
\footnotesize
\begin{tabularx}{\linewidth}{@{}lX@{}}
\toprule
Layer & Selectable features \\
\midrule
Frame & numerology 0--2, bandwidth, TDD pattern and special slot, RBG size \\
Radio & log-distance or TR~38.901~\cite{3gpp38901} path loss, radio maps, correlated shadowing, per-subband Rayleigh fading with per-robot Doppler, thermal or fixed noise, blockage \\
PHY & MCS table 1 or 2, exact TB size, EESM, Sionna or 5G-LENA BLER tables, MCS caps \\
Access & SR period, fixed access delay or buffer-report grants, proactive grants, processing offset \\
Scheduler & PF, maximum rate, or round robin, PF update per slot or per RBG, retransmissions sharing a slot or taking a whole slot \\
HARQ/RLC & processes, maximum transmissions, chase or incremental-redundancy combining, RLC AM retry or UM loss, Packet Data Convergence Protocol (PDCP) discard \\
Downlink & gNB queues, CQI period and delay, HARQ feedback timing \\
Cells & 1--7 cells, uplink power control, same-slot interference, A3 handover \\
Application & message sizes, traffic models, FIFO depth, timeout, control step \\
\bottomrule
\end{tabularx}
\end{table}

Presets bundle these settings for common scenarios.
The \texttt{lena\_validation} preset, called v1 below, reproduces the ns-3 5G-LENA scenario of Section~\ref{sec:eval-lena} with one fitted access delay, and \texttt{lena\_validation\_v2}, called v2, adds the four load options with no fitted value.
The \texttt{srsran\_like} and \texttt{oai\_like} presets carry uplink delay parameters fitted on the prototype engine to public measurements of the srsRAN~\cite{gomezmiguelez2016srslte} and OAI~\cite{nikaein2014oai} stacks.
The \texttt{oai\_rfsim} preset carries three values measured on a running OAI stack, namely the SR period, the delay from a grant to the data it allows, called K2, and the HARQ round-trip time.
It adds one access delay fitted to the OAI traces, as reported in the later Section~\ref{sec:eval-calib}.
Each level declares which field groups it reads, and a strict mode rejects a configuration that sets a field the level cannot honor, such as a path-loss exponent under the i.i.d. delay model.

\subsection{Multiple Cells}
\label{sec:multicell}

A fleet that spans a warehouse is typically served by several gNBs, so \alg supports up to $C$ (e.g., $7$) cells per environment in hexagonal, grid, or custom layouts, batched over a cell dimension.
Each cell runs its own PF scheduler over the carrier that all cells share.
Uplink interference at a gNB comes from the robots that other cells scheduled on the same subband in the same slot, each through its own fading.
Link adaptation uses the interference of the previous slot, as a real gNB would, while decoding uses that of the current slot, and OLLA absorbs the resulting error.
Fractional uplink power control is on by default with more than one cell, because without it three cells can deliver less traffic than a single cell.
This scheme compensates only part of each robot's path loss, which limits the interference that robots cause in other cells.
Handover follows the 3GPP A3 event, which fires when a neighbor cell stays stronger than the serving cell by a hysteresis margin for a time-to-trigger.
The multi-cell model runs on the reference and graph backends of the NR engine and on the reference backend of the prototype engine.
With one cell, it is bitwise identical to the single-cell engine.

\subsection{Coupling with Isaac Lab}
\label{sec:isaac}

Isaac Lab runs physics substeps of length $\delta$ (e.g., $20$~ms) and applies the policy every $D$ (e.g., $5$) substeps, so one control step lasts $\Delta = D\delta$.
\alg aligns its $K$ uplink slots with exactly this interval.
A mixin class couples \alg with direct-workflow environments through four hooks for setup, stepping, reset, and observations.
In its pre-physics hook, the environment decides what each robot sends, and \alg captures those messages at the start-of-step pose.
Thus, a message never reports a state that the robot has not reached yet.
The network step runs after the physics and before Isaac Lab resets finished environments.
The reset hook forwards the indices of these environments to the partial reset of the engine.
The observation hook returns, per robot, the AoI of its freshest delivered message, its SNR, its queued messages, and whether its last message was delivered.
A standard event term redraws per-environment radio parameters at every reset, thereby enabling domain randomization of the network.

\section{Implementation}
\label{sec:impl}

\myparatight{Package Structure}
\alg ships as the pip-installable Python package \texttt{isaac-net} at \url{https://github.com/ZzZTripleZzZ/isaac-net}, which depends only on PyTorch and NumPy and imports Triton lazily for the triton backend.
A backend-agnostic core holds the engines, the configuration, and the radio, and a single factory, \texttt{make\_engine}, builds every fidelity level with the same dictionary of outputs.
Each engine runs on the reference backend, on the graph backend, which is bitwise identical to it, or on the triton backend, which matches it to floating-point rounding but lacks the buffer-report grant procedure.
A thin Isaac Lab integration layer builds its engine through \texttt{make\_engine} and adds the environment mixin, its own radio model with blockage and domain randomization, and the event terms.
The package ships under the BSD-3-Clause license with BLER tables exported from Sionna 2.2.0 after correcting defects in its table data, such as curves stored under the wrong MCS, while the GPL-licensed 5G-LENA tables are generated locally.
The same network module also runs inside MuJoCo Playground through a JAX callback, and its outputs there are bitwise equal to a direct PyTorch replay of the same poses.

\myparatight{Isaac Lab Integration}
The Isaac Lab integration layer was validated on Isaac Lab 3.0 with Isaac Sim 6.1 on Windows 11, and on Linux without Isaac Sim using the Newton and PhysX physics backends.
The example environment implements a fleet task in which $R$ spheres driven by velocity commands move in a 150~m square arena.
The fleet learns about a randomly appearing hazard only when the uplink delivers a large message, standing in for a camera frame, from a robot that observed it.
An end-to-end Proximal Policy Optimization run with the rsl\_rl library trains this task with the network on the triton backend in the loop at $1024 \times 16$ for 30 iterations, which checks the training pipeline end to end.

\myparatight{Test Suite}
The test suite checks the MAC timeline slot by slot and checks TB sizes against TS~38.214~\cite{3gpp38214} examples and 564,300 cases computed by Sionna.
It also runs the equivalence checks of Section~\ref{sec:backends} on the GPU and verifies that the network inside a running Isaac Lab environment equals a replay of its poses and sends on the reference backend across partial resets.
Continuous integration runs the CPU tests on every push, 690 CPU tests pass against the installed release package, and the last full GPU run passed all 75 GPU tests.

\myparatight{Co-Simulation Bridges}
For validation, \alg includes bridges that run the same task against ns-3 5G-LENA~\cite{patriciello2019lena} in lockstep every control step, as a process pool with one ns-3 process per environment, or offline on a recorded rollout.
Each bridge reproduces a standalone 5G-LENA run of the same scenario message by message, except six messages in one run that waited about 10~s and were purged with the bridge's message registry.
A propagation filter stops ns-3 from propagating each robot's uplink signal to every other robot, which no single-cell receiver uses but which costs time quadratic in $R$, without changing any delivery time.
A further bridge runs the full OAI 5G stack~\cite{nikaein2014oai} with its core network and RF simulator, one UE per robot, in lockstep with the control loop for up to ten robots.

\myparatight{Current Limits}
Each triton kernel processes all robots of one environment in one GPU program, so beyond about 128 robots per environment for the NR engine, and 256 for the prototype engine, it would need to split the scheduler's search over robots into blocks.
The triton kernel of the NR engine already spills registers, that is, moves per-robot state to slower memory, at 64--100 robots, so porting the nine per-robot arrays of the buffer-report grant procedure remains a follow-up.

\section{Evaluation Results}
\label{sec:eval}

\subsection{Experimental Setup}
\label{sec:eval-setup}

\myparatight{Testbed and timing protocol}
All GPU timings ran on an NVIDIA RTX 4090 with an Intel Core i9-13900KF host, and a dedicated NVIDIA L40 of an HPC cluster served as a second GPU.
Before every case, the benchmark driver waited until the GPU was idle.
In the first campaign, all 516 processes that timed the network alone or the fleet task and all 9 Isaac Lab runs started at 0\% utilization with 36~MiB in use.
A follow-up campaign on the same day timed the scale configuration of Section~\ref{sec:eval-lena}, starting at 0\% utilization with 1,187~MiB held by an idle desktop session.
The host CPU was not reserved, with one single-threaded job of another user in the first campaign and Windows Defender using up to 4 cores, with a median of 2.3, in the follow-up.
Each case that timed the network alone ran in three fresh processes that each timed three windows of about 2~s after warm-up, and we report the median over processes of each process's median window.
In the follow-up, each Isaac Lab size ran three processes with the network off and three with the NR engine, interleaved, and we report the mean and range of their medians over three windows of 50 control steps.
The prototype-engine Isaac Lab runs of the first campaign used one process each.
We compare against ns-3.48 with 5G-LENA v5.1~\cite{patriciello2019lena,koutlia2022lenacalib}, patched only against an abort in its uplink channel-quality path.
Its timings come from the 32 hardware threads of the same machine while other jobs shared it, so they are possibly 2--3$\times$ pessimistic.

\myparatight{Default parameters}
Unless otherwise stated, a control step lasts $\Delta = 100$~ms and covers $K = 40$ uplink slots at 30~kHz subcarrier spacing in a DDDSU pattern.
This pattern holds three downlink slots, one special slot, and one uplink slot per 2.5~ms.
The NR engine uses a 20~MHz carrier with 13 RBGs and $H = 16$ HARQ processes, except that the 5G-LENA presets adopt the 50 PRBs in 5 RBGs of the 5G-LENA scenario, and the prototype engine uses $S = 5$ subbands of 10 PRBs each.
Each robot has a FIFO of $F = 16$ messages with a 2~s timeout, the message classes carry 4,000 and 30,000 bytes, and the radio averages $J = 4$ poses per step.
Isaac Lab runs substeps of $\delta = 20$~ms with $D = 5$ per control step, up to $C = 7$ cells are supported, and sizes $E \times R$ range from $8 \times 16$ to $8192 \times 128$.
The 5G-LENA sweep uses $N \in \{1, 2, 4, 8, 16, 32, 64\}$ UEs and messages of $B \in \{4000, 30000\}$ bytes.
When the network is timed alone, every robot sends with probability 0.3 per step, a third of them large messages, which saturates the uplink from $R = 16$ on.

\subsection{Equivalence of the Fast Backends}
\label{sec:eval-equiv}

We run the equivalence tests of Section~\ref{sec:backends} for 300 control steps with a workload that cycles through idle, medium, and saturating phases and lets the SNR drift over $[-10, 40]$~dB.
For the NR engine, the workload also applies 22--42 random partial resets per run.
We observe that the graph backend of the NR engine is bitwise identical to the reference in every output and state tensor at every step in 13 runs over ten configurations.
These configurations include the downlink, three cells with handover, and the 5G-LENA settings.
The 13 runs used code from before the final merges, and shorter runs of every check pass on the final code.
In the per-step mode, which starts both backends from the same recorded state at every step, the NR triton backend disagrees with the reference in 3 of 218,442 robot-steps with traffic at $64 \times 16$ and 35 of 1,910,974 at $256 \times 32$, that is, about 1.4 and 1.8 per 100,000.
In the scale configuration, it disagrees in 2 of 262,080 robot-steps at $64 \times 16$.
This is because the triton kernel reduces in a different order, so a value at a threshold, such as an MCS boundary, occasionally flips.
In the full-run mode, such a flip can change the future of one robot, yet over 300 steps at $64 \times 16$ the two backends deliver 59,720 versus 59,710 messages with a mean delay of 7.6147 versus 7.6145 steps.
The prototype engine behaves similarly, with a bitwise identical graph backend and a triton backend that disagrees in 5 of 1,631,514 robot-steps at $64 \times 100$.
In contrast, \texttt{torch.compile} changes completion times in 11.7\% of its robot-steps with traffic, because its rounding leaves byte leftovers large enough to delay fully served messages, which the earlier Eq.~\eqref{eq:fifo} prevents.
Thus, graph runs are bitwise identical to the reference, while triton runs, which carry every scale result of Section~\ref{sec:eval-isaac}, match it in aggregate.

\subsection{Fidelity Against ns-3 5G-LENA}
\label{sec:eval-lena}

\myparatight{5G-LENA scenario}
The 5G-LENA scenario simulates a single-cell uplink in which one gNB serves $N$ UEs in a 100~m square arena on a 3.5~GHz, 20~MHz carrier with 50 PRBs in 5 RBGs.
It uses path loss of $40 + 35\log_{10} d$, and 6~dB shadowing per link.
The scenario runs the default PF scheduler, SR and BSR, HARQ with at most four transmissions, the EESM error model, and RLC in UM with a 2~s discard timer.
Every 100~ms, each UE sends a message of $B$ bytes with probability $p$.
By default, 5G-LENA selects the uplink MCS without OLLA from the SINR of the UE's last transmission, so with 3GPP fading it drops a median 70\% of messages, which makes such runs unusable for comparison.
We therefore disable fading, spread UE power over the whole band, and redraw any UE placed where even the lowest MCS fails.
The primary sweep combines every $N$ and $B$ value of Section~\ref{sec:eval-setup} with nominal loads from 10\% to 120\% of the saturated throughput and three seeds of 30~s, for 153 runs.
The NR engine receives an option for each mismatch that it can express, such as a wideband PF metric, no OLLA, the 5G-LENA BLER tables, and loss of a TB after four failed transmissions.
It replays each run with four engine replicas that differ only in their random draws, using the per-UE signal strengths and the exact message times that 5G-LENA exports.
We compare delay quantiles, drop rates, PRB use, and the Kolmogorov--Smirnov (KS) distance, the largest gap between the two delay CDFs, and the Wasserstein-1 (W1) distance, the average horizontal gap between them.
A run has light, moderate, or saturated load when 5G-LENA drops under 1\%, 1--20\%, or at least 20\% of messages.
For each load, Table~\ref{tab:lena} gives the median over runs of the signed and absolute relative errors of the per-run p50 delay, the relative p95 error, the drop-rate difference, the KS distance, and the relative error of PRB use.
It compares four configurations, namely the default \texttt{NRConfig()}, v1 with one fitted access delay, v2 with the four load options and no fitted value, and the scale configuration, labeled v2 w/o BSR.
Default and v2 w/o BSR were replayed on the triton backend and v1 and v2 on the reference backend, which agree to within 0.8 points and 0.003 in KS on check runs.

\begin{table}[t]
\centering
\caption{NR engine configurations vs. 5G-LENA by load, as the median error over runs (closer to zero is better).}
\label{tab:lena}
\footnotesize
\setlength{\tabcolsep}{2.1pt}
\begin{tabular}{@{}llrrrrrr@{}}
\toprule
Load (runs) & Configuration & p50 & $|$p50$|$ & p95 & Drop & KS & PRB \\
 & & (\%) & (\%) & (\%) & (pp) & & (\%) \\
\midrule
Light (79) & Default & $-30.2$ & 30.2 & $-47.1$ & $0.00$ & 0.61 & $-37.4$ \\
 & v1 & $-0.2$ & 4.2 & $-9.2$ & $0.00$ & 0.26 & $-13.4$ \\
 & v2 w/o BSR & $-3.5$ & 4.5 & $-8.9$ & $0.00$ & 0.24 & $-12.9$ \\
 & v2 & $-0.2$ & 2.5 & $-2.5$ & $0.00$ & 0.25 & $-3.2$ \\
\midrule
Moderate (36) & Default & $-43.8$ & 44.1 & $-73.7$ & $-5.42$ & 0.62 & $-36.5$ \\
 & v1 & $-26.9$ & 26.9 & $-15.4$ & $-4.05$ & 0.32 & $-5.4$ \\
 & v2 w/o BSR & $-5.7$ & 6.6 & $-14.5$ & $-2.44$ & 0.14 & $-6.3$ \\
 & v2 & $-1.2$ & 4.8 & $-6.1$ & $-1.42$ & 0.13 & $-4.9$ \\
\midrule
Saturated (38) & Default & $-76.3$ & 76.9 & $-75.8$ & $-37.92$ & 0.77 & $-8.1$ \\
 & v1 & $-33.7$ & 34.8 & $-3.3$ & $-6.84$ & 0.38 & $+0.0$ \\
 & v2 w/o BSR & $-0.8$ & 5.5 & $-0.2$ & $-2.57$ & 0.08 & $-3.1$ \\
 & v2 & $-0.1$ & 4.4 & $+0.1$ & $-1.70$ & 0.09 & $-3.0$ \\
\bottomrule
\end{tabular}
\end{table}

\myparatight{One fitted access delay matches light load only}
The \texttt{lena\_validation} preset, v1 in Table~\ref{tab:lena}, fits one parameter on seeds 1 and 3, the delay from an SR to the first uplink data.
It selects 20~ms, that is, 40 consecutive slots of 0.5~ms counted over all slot types, not the $K$ uplink slots.
On the 51 runs of the held-out seed, most of them light, its median p50 and p95 errors are $-0.2$\% and $-6.8$\%.
However, 5G-LENA traces show only about 4.5~ms from a request to the next grant and about 7~ms to the data, so the fitted 20~ms absorbs other delays that the buffer-report grant procedure models explicitly.
Moreover, v1 is optimistic under load, where its median delay is about 27\% and 34\% low at moderate and saturated load, respectively.

\begin{figure}[t]
\centering
\includegraphics[width=\columnwidth]{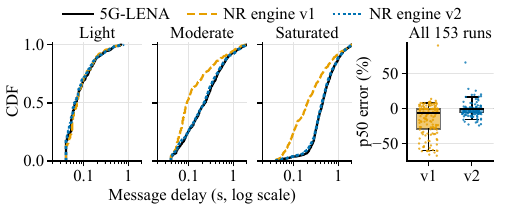}
\caption{Per-message delay CDFs of 5G-LENA and the NR engine v1 and v2 in three runs, with the p50 delay error of all 153 runs as a box plot.}
\label{fig:lena-cdf}
\end{figure}

\myparatight{Modeling the load behavior removes the delay bias}
The v2 preset turns on the four options of Section~\ref{sec:nrconfig} and has no fitted parameter, since it reads every timing value of the buffer-report grant procedure from the configuration and traces of 5G-LENA.
We observe that v2 brings the median p50 error to $-1.2$\% and $-0.1$\% at moderate and saturated load, respectively, and keeps it at $-0.2$\% at light load.
The median absolute error of the per-run median delay is 2.5\%, 4.8\%, and 4.4\% at the three loads, so the small signed medians do not hide a symmetric spread.
Over all 153 runs, the median KS distance falls from 0.28 to 0.17 and the median absolute p95 error from 8.6\% to 3.5\%.
Figure~\ref{fig:lena-cdf} shows one light, one moderate, and one saturated run with $N = 32$, $B = 30{,}000$ bytes, and $p = 0.031$, $0.063$, and $0.094$ on the held-out seed.
At every load, the dotted v2 curves follow the solid 5G-LENA curves, while the dashed v1 curves lie to their left under load.
Its box plot gives the relative p50 error of all 153 runs, with quartile boxes and whiskers at the 5th and 95th percentiles.
v2 follows 5G-LENA under load because both update a UE's PF average after every RBG and freeze the averages of idle UEs, so backlogged UEs share every slot, whereas v1 serves messages nearly one at a time.
Furthermore, nearly every message whose UE empties its buffer leaves 5--17 bytes of RLC header that are reported only after a 10~ms status timer, so this wait adds about 15~ms, most of the 20~ms that v1 fitted.
Two residuals remain: 1) v2 still drops 1.4--1.7 percentage points too few messages under load, possibly due to a link-adaptation failure of 5G-LENA for a few cell-edge UEs; and 2) its first-transmission BLER exceeds that of 5G-LENA by 0.5 points, while its uplink MCS is one step above that of 5G-LENA for 49\% of 3,583 UEs.
The two fit together, since a link adaptation that runs one MCS higher fails slightly more first transmissions.
At light load, the KS distance stays near 0.25 because both delay distributions are only a few milliseconds wide, so a one-millisecond shift moves a large share of the CDF.

\myparatight{The scale configuration stays close to v2}
The triton kernel supports every v2 option except the buffer-report grant procedure, so the scale experiments of Section~\ref{sec:eval-isaac} use v2 with the single fixed 40-slot access delay of v1 in its place, which forms the scale configuration, labeled v2 w/o BSR in the tables.
Its median absolute p50 error of 4.5--6.6\% and its KS distances stay close to those of v2.
Without the procedure, the tail and the light-load PRB use move back toward v1, and the drop gap under load grows by about one percentage point.
A FIFO of 16 messages, as in the scale experiments, changes only the saturated drop gap, to $-2.9$ points.
In contrast, the default configuration, \texttt{NRConfig()}, differs from v2 in the access delay, the BLER tables, OLLA, and the scheduler and retransmission rules, and its median delay is 30\%, 44\%, and 76\% low at the three loads.
Section~\ref{sec:eval-isaac} therefore times the scale configuration, which stays close to 5G-LENA.

\myparatight{A held-out carrier}
The four v2 options and an 8-byte header overhead per TB were chosen while studying the 153 runs, so we also replay one 5G-LENA scenario that none of these choices was checked against.
It uses a 10~MHz carrier with 20 PRBs in 2 RBGs, UE placements from a fourth seed, and $N \in \{8, 16, 32\}$, for 39 runs, while v2 changes only its carrier fields.
Under moderate and saturated load, the delay distributions stay as close as in the sweep, with KS distances of 0.14 and 0.08 and p95 errors of $-7.8$\% and 0.0\% in the 6 moderate runs and the saturated runs, respectively.
However, the median delay is 5--10\% low at every load, and at saturated load the engine drops 4.1 percentage points too few messages, against 1.7 in the sweep.
The largest errors come from the two saturated runs with 16 UEs and 4,000-byte messages, where the median delay of the engine is 125~ms against 229 and 321~ms in 5G-LENA, that is, 45\% and 61\% low.
Both residuals of the sweep reappear in the same direction, as the first-transmission BLER is 0.7 points high and the engine picks a higher MCS for 55\% of the UEs.
One MCS step is a larger share of the capacity of a 20-PRB cell, which possibly explains the low median.
The closed-loop runs of Section~\ref{sec:eval-closed} show an offset of the opposite sign, so the out-of-sample error of the median delay has no fixed direction.

\myparatight{Link tables and the prototype engine}
Interestingly, at the same MCS, the 5G-LENA tables need 2.4 to 5.7~dB more SINR than the Sionna tables to reach a BLER of 10\%.
Replaying with the Sionna tables therefore leaves the p95 delay and PRB use 32\% and 33\% low, respectively.
Thus, a comparison across tables would measure the tables, not the MAC.
The prototype engine replays the same runs with a median absolute p95 error of 36\%, against 8.6\% for v1.
At saturated load, it drops 32 percentage points fewer messages than 5G-LENA, likely because its logistic PHY is optimistic at high spectral efficiency.
Lastly, the validation covers the 100~m arena only, since 24\% of positions in the 150~m arena of the fleet task lie below the SNR range that the 5G-LENA runs covered.
In those positions, 5G-LENA delivers only 0--4\% of what the prototype engine delivers.
These results underscore that a per-slot GPU engine closely reproduces a packet-level simulator from light load to saturation once it models the scheduler and the buffer-report grant procedure.

\subsection{Calibration Against Real 5G Stacks}
\label{sec:eval-calib}

\myparatight{Public measurements anchor individual layers}
We fit the access procedure of the prototype engine to one-way uplink delays of the OAI and srsRAN~\cite{gomezmiguelez2016srslte} stacks from a 5G campus dataset~\cite{raffeck2024campusdata}.
The fit reduces the W1 distance from 2.8 to 1.25~ms for srsRAN and from 3.3 to 2.5~ms for OAI.
The fitted procedure reaches 2.0--3.1~ms on 36 held-out configurations, and this fit was not repeated on the NR engine.
Notably, a single fixed access delay fits neither stack, whereas a cycle of scheduling requests and grants of about 20~ms likely explains srsRAN, and OAI likely grants uplink resources before a request arrives, which fits its short TDD periods.
The uncalibrated prototype engine overestimates capacity by about 25--35\% on the ColO-RAN dataset~\cite{polese2023coloran}.
Its 6~dB shadowing lies inside the 4--8~dB of two outdoor drive tests on the POWDER testbed~\cite{nawazish2026powderho,kamath2026powderviavi} but its shadowing changes with distance two to four times faster than measured.

\myparatight{A running 5G stack corrects the access presets}
Because public traces do not show where the access time goes, we also ran OAI 2026.w39 with its 5G core in RF-simulator mode on the 5G-LENA frame structure.
In this mode, every packet passes through the real PDCP, RLC, MAC, scheduler, and HARQ code, while the radio channel is simulated.
Its traces show an SR period of 5~ms, a K2 of 6 slots, and a HARQ round-trip time of 10 slots.
In these traces, a 100-byte message waits a median 14.1~ms before its first uplink data.
These values contradicted the SR period, access delay, K2, and HARQ round-trip time of both the \texttt{oai\_like} and 5G-LENA presets.
The \texttt{oai\_rfsim} preset takes the measured SR period, K2, and HARQ round-trip time and fits one access delay to the traces by replay.
Replayed through the NR engine, the \texttt{oai\_rfsim} preset matches single-UE delays of small messages to a median W1 distance of 2.5~ms, against 13.3~ms for \texttt{oai\_like} and 13.2~ms for the 5G-LENA preset.
On held-out runs, its median W1 distance is 5.5~ms.
Yet no preset reproduces OAI saturation at 4.8~Mbit/s of offered load or its multi-UE tails.
The p99 delay of OAI grows from 61~ms at one UE to 483~ms at ten, in a run at 0.05$\times$ real time on a loaded host, possibly because its gNB has no room left in the control channel to send grants to every UE.

\subsection{Speed of the Network Engines}
\label{sec:eval-speed}

\begin{table}[t]
\centering
\caption{Network step time in ms per 100~ms control step on an idle RTX 4090 (lower is better).}
\label{tab:speed}
\footnotesize
\setlength{\tabcolsep}{2.2pt}
\begin{tabular}{@{}lrrrrrr@{}}
\toprule
& \multicolumn{3}{c}{$256 \times 16$} & \multicolumn{3}{c}{$4096 \times 100$} \\
\cmidrule(lr){2-4}\cmidrule(lr){5-7}
Level & ref. & graph & triton & ref. & graph & triton \\
\midrule
L0 & 1.23 & 0.36 & -- & 4.75 & 10.6 & -- \\
L1 & 15.1 & 1.94 & 0.36 & 107 & 108 & 11.1 \\
L2-legacy & 94.0 & 9.71 & 0.46 & 191 & 135 & 17.7 \\
L2, UL & 330 & 50.2 & 1.68 & 2,310 & 2,228 & 67.8 \\
L2, UL+DL & 1,540 & 240 & 5.66 & 11,467 & 11,084 & 295 \\
L2, v2 w/o BSR & -- & -- & 1.97 & -- & -- & 94.4 \\
\bottomrule
\end{tabular}
\end{table}

\myparatight{The triton kernel is the path to scale}
In Table~\ref{tab:speed}, the L2 rows run the default configuration, UL+DL adds the downlink (DL) with 160 data slots per step to the uplink (UL), and v2 w/o BSR, the scale configuration, was timed on the triton backend only.
We observe that the triton backend advances $4096 \times 100$, that is, 409,600 robots, by one control step of the NR uplink in 67.8~ms.
In contrast, the reference and graph backends require 2,310~ms and 2,228~ms, respectively.
This is because the triton kernel launches once per control step and keeps each environment's state in registers, whereas at this size the other two backends are bound by the memory traffic of tens of thousands of small kernels.
The graph backend is 6.6$\times$ faster than the reference at $256 \times 16$ and gives runs that are bitwise identical to the reference.
On the triton backend, the NR uplink requires approximately 3.7--3.8$\times$ the time of the prototype engine at both sizes, likely due to its 29 MCS levels, the EESM of every candidate MCS, and its 16 HARQ processes.
The scale configuration costs 17--39\% more than the default on the triton backend, at 1.97 against 1.68~ms and 94.4 against 67.8~ms, perhaps because per-RBG PF updates and link adaptation based on the previous allocation add work to every slot.
On the triton backend, the downlink multiplies the cost of the default uplink by 3.4--4.3$\times$.
The i.i.d. delay and fluid models cost at most 11.1~ms per step at $4096 \times 100$.
There, the graph backend of the i.i.d. delay model is slower than the reference, because its fixed-shape removal of finished messages rewrites every queue entry, whereas the plain PyTorch reference touches only new and finished messages.
On the dedicated L40, the same triton kernel of the NR engine requires 72.6~ms at $4096 \times 100$, within 8\% of the RTX 4090.
Thus, the triton kernel advances the NR uplink of 409,600 robots in less than one 100~ms control step in both configurations.
Inside Isaac Lab, the network alone takes about a fifth to a quarter of each environment step from 262,144 to 1,048,576 robots, as the following Section~\ref{sec:eval-isaac} shows.

\subsection{Scale Inside Isaac Lab}
\label{sec:eval-isaac}

\begin{table}[t]
\centering
\caption{Isaac Lab fleet task at 128 robots per environment with the network off and with the NR engine in the scale configuration, as the mean over three processes with the range below.}
\label{tab:isaac}
\footnotesize
\setlength{\tabcolsep}{2.5pt}
\begin{tabular}{@{}rrrrrr@{}}
\toprule
& Off & L2 & Added & \multicolumn{2}{c}{Network alone (ms)} \\
\cmidrule(lr){5-6}
Robots & st/s & st/s & (ms) & L2 & L2-legacy \\
\midrule
262,144 & \makecell[r]{5.38\\[-2pt]{\scriptsize 5.04--5.61}} & \makecell[r]{4.11\\[-2pt]{\scriptsize 3.82--4.34}} & \makecell[r]{57\\[-2pt]{\scriptsize 52--63}} & 50 & 10.9 \\
524,288 & \makecell[r]{2.95\\[-2pt]{\scriptsize 2.84--3.07}} & \makecell[r]{2.44\\[-2pt]{\scriptsize 2.37--2.52}} & \makecell[r]{71\\[-2pt]{\scriptsize 45--86}} & 100 & 22.9 \\
1,048,576 & \makecell[r]{1.50\\[-2pt]{\scriptsize 1.37--1.57}} & \makecell[r]{1.24\\[-2pt]{\scriptsize 1.23--1.24}} & \makecell[r]{140\\[-2pt]{\scriptsize 80--171}} & 199 & 45.2 \\
\bottomrule
\end{tabular}
\end{table}

\myparatight{One million robots with the network in the loop on one GPU}
We run the fleet task of Section~\ref{sec:impl} under a random policy, that is, with random actions, under which each robot sends on about two thirds of the steps, half of them large messages.
The uplink therefore saturates, which is the worst case for network cost.
The NR engine runs the scale configuration of Table~\ref{tab:lena} with the 16-message FIFO of the task, and each robot's SNR comes from the engine's radio with thermal noise and no fading.
The fleet arena serves here to measure cost, and the fidelity of this configuration rests on Table~\ref{tab:lena}.
At $R = 128$, the triton kernel of the NR engine runs at the largest size per environment that it supports without splitting its scheduler search into blocks (Section~\ref{sec:impl}).
Table~\ref{tab:isaac} gives control steps per second (st/s, higher is better), the step-time increase of each adjacent off and on pair, and the time of the network alone per step, with the prototype engine from one earlier process.
We observe that $8192 \times 128$, or 1,048,576 robots, runs at 1.24 control steps per second with the NR engine in the loop against 1.50 without the network.
This rate equals 1.30 million robot-steps per second within 19.5~GiB of device memory, of which the idle desktop session holds 1.2~GiB.
It covers the environment step that a training loop calls, including physics, network, and observations, but not the policy forward pass or the learning update.
Alone, the NR engine requires 50, 100, and 199~ms per control step at the three sizes, about a fifth to a quarter of the whole step and four to five times the prototype engine.

\begin{figure}[t]
\centering
\includegraphics[width=\columnwidth]{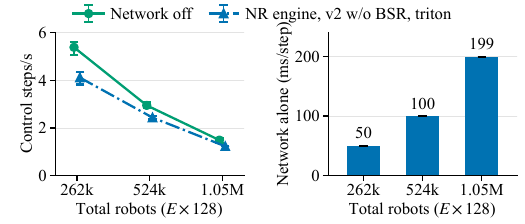}
\caption{Control steps per second and network time per control step for Isaac Lab with $E \times 128$ robots, as the mean and range over three processes (higher is better on the left, lower on the right).}
\label{fig:isaac-scale}
\end{figure}

\myparatight{Part of the network hides behind host work}
Figure~\ref{fig:isaac-scale} plots the rates on the left and the time of the network alone on the right, and from 524,288 robots on, the rate with the network moves closer to the rate without it, which reflects the overlap discussed next.
At 262,144 robots, the step grows by 52--63~ms, at least the cost of the network alone, so nothing overlaps there, possibly because the Isaac Lab layer adds CPU-side work of its own for the radio and the observations.
At 524,288 and 1,048,576 robots, the step grows by 45--86 and 80--171~ms, respectively, that is, by 40--86\% of the cost of the network alone.
Thus, from 524,288 robots on, the median pair of network-off and network-on processes hides 18--19\% of the network's GPU time while the CPU runs PhysX and Python work.
This overlap is possible because the Isaac Lab step keeps the device-wide GPU utilization at 22--38\%.
The mean added time of Table~\ref{tab:isaac} suggests about 30\% instead, because one slow network-off process per size lowers it.
Scene startup, not stepping, limits scale, as it grows roughly linearly with the number of robots to 1,158--1,585~s at one million robots.
We take two control steps per second as a practical bar for on-policy training, which learns only from experience collected by the current policy.
By this bar, the scale configuration keeps the environment step fast enough for training up to 524,288 robots on one GPU and runs one million robots at 83\% of the rate without the network.

\subsection{Cost of Packet-Level Co-Simulation}
\label{sec:eval-ns3}

\myparatight{ns-3 cost grows with robots, not with load}
We measure what massively parallel training would cost if the ns-3 process pool of Section~\ref{sec:impl} replaced the GPU engine, which is a cost of fidelity and not a like-for-like speed-up, since 5G-LENA simulates far more detail.
One ns-3 process requires 15 to 183~ms per control step from 8 to 100 robots with the propagation filter of Section~\ref{sec:impl}, about $R^{0.91}$, and 25 to 1,527~ms without it, about $R^{1.57}$.
Offered load has little effect, and with 12 workers the pool reaches about 2,100--2,900 robot-steps per second at every $R$, a parallel efficiency of 0.30--0.48.

\myparatight{Co-simulation cannot keep pace with GPU training}
At $4096 \times 128$, ns-3 would require 1,071 CPU-seconds per control step, with the per-worker cost extrapolated from $R = 100$ to 128 by the fitted cost model.
On 32 threads at perfect parallel efficiency, this cost falls to 33.5~s, plus the Isaac Lab step without the network.
With the network on the GPU, Isaac Lab requires 0.41~s per step with the NR engine in the scale configuration, the mean of three processes between 0.40 and 0.42~s, which is about 83$\times$ faster.
With the prototype engine in one process, it requires 0.42~s, about 81$\times$ faster, and the two engines lie within the process-to-process spread, so these times do not rank them.
In the units of Fig.~\ref{fig:teaser}, this training loop would reach about 15,500 robot-steps per second with ns-3 against about 1.28 million with the NR engine.
For the network alone, the NR engine requires 99.8~ms per step inside Isaac Lab, 340$\times$ less than ns-3 on 32 threads, and the prototype engine 22.9~ms, 1,460$\times$ less.
Both ratios assume a parallel efficiency that the pool never reached, which roughly offsets the effect of the shared machine noted in Section~\ref{sec:eval-setup}.
Wall-clock co-simulation is slower still, as ns-3's real-time scheduler held wall-clock pace reliably for only about two robots and never at 16 or more.
One ns-3 process of the pool, run in lockstep and paced at 10~Hz, does better, with a per-step compute under 100~ms at the 99th percentile up to 48 robots in one cell, that is, 480 robot-steps per second, as Fig.~\ref{fig:teaser} shows.
Thus, once the network runs on the GPU, the physics dominates the environment step, while packet-level simulators suit validation at small scale.

\subsection{Multi-Cell Behavior}
\label{sec:eval-multicell}

In a sweep of 16 environments of 32 random-walking robots over 150 steps with identical poses and traffic, the NR and prototype engines agree to about 1~dB in SINR without power control.
With power control, which is on by default, the two engines differ by about a third in delivered bytes, with 160 against 117~kB per environment and control step for three cells at 8.4~kB per robot-step, and neither is calibrated.
The NR engine then delivers 44, 76, and 185~kB per environment and control step with one, three, and seven cells at 15~kB per robot-step.
When one robot drives through three cells in 32 environments, two boundary crossings trigger exactly two handovers in 30 of them.
The robot is first served in the first uplink slot after the 40~ms handover interruption of the default multi-cell preset.
Thus, the multi-cell timing behaves as specified.

\subsection{Closed-Loop Behavior of the Fidelity Levels}
\label{sec:eval-closed}

\begin{table}[t]
\centering
\caption{One scripted controller in closed loop over each network model in three traffic settings, as the mean over five seeds, with the KS distance of the delays to ns-3 (lower is closer).}
\label{tab:closed}
\footnotesize
\setlength{\tabcolsep}{2.4pt}
\begin{tabular}{@{}lrrrrr@{}}
\toprule
Network & p50 (ms) & p95 (ms) & AoI p95 (s) & Sent & KS \\
\midrule
\multicolumn{6}{@{}l}{\textit{Light load, one message per 0.6~s, $8 \times 16$}} \\
Ideal link & 0 & 0 & 0.60 & 6,400 & \\
L0, fitted to L2 & 255 & 372 & 0.80 & 6,394 & 0.10 \\
L0-emp, L2 resampled & 255 & 415 & 0.90 & 6,394 & 0.11 \\
L1 & 270 & 289 & 0.80 & 6,400 & 0.58 \\
L2-legacy & 279 & 320 & 0.80 & 6,400 & 0.46 \\
L2, v2 preset & 257 & 409 & 0.86 & 6,393 & 0.11 \\
ns-3 5G-LENA & 252 & 399 & 0.90 & 6,386 & \\
Seed noise & & & & & 0.08--0.10 \\
\midrule
\multicolumn{6}{@{}l}{\textit{One message per 0.2~s, $8 \times 16$}} \\
Ideal link & 0 & 0 & 0.20 & 19,200 & \\
L0, fitted to L2 & 259 & 404 & 0.70 & 12,083 & 0.22 \\
L0-emp, L2 resampled & 260 & 507 & 0.80 & 11,042 & 0.11 \\
L1 & 270 & 285 & 0.50 & 12,772 & 0.50 \\
L2-legacy & 259 & 327 & 0.54 & 12,671 & 0.28 \\
L2, v2 preset & 263 & 506 & 0.96 & 11,062 & 0.12 \\
ns-3 5G-LENA & 254 & 500 & 0.98 & 11,024 & \\
Seed noise & & & & & 0.12--0.13 \\
\midrule
\multicolumn{6}{@{}l}{\textit{One message per 0.6~s, $8 \times 32$}} \\
Ideal link & 0 & 0 & 0.60 & 12,800 & \\
L0, fitted to L2 & 524 & 811 & 1.30 & 11,864 & 0.11 \\
L0-emp, L2 resampled & 525 & 1,027 & 1.62 & 11,284 & 0.35 \\
L1 & 540 & 565 & 1.10 & 12,797 & 0.58 \\
L2-legacy & 458 & 586 & 1.06 & 12,720 & 0.28 \\
L2, v2 preset & 525 & 1,039 & 2.00 & 11,235 & 0.35 \\
ns-3 5G-LENA & 483 & 979 & 1.94 & 11,353 & \\
Seed noise & & & & & 0.05--0.06 \\
\bottomrule
\end{tabular}
\end{table}

\myparatight{Setup}
We run one scripted controller of the fleet task over each network model at $8 \times 16$ for 300 steps and five seeds, with ns-3 in lockstep.
Each robot sends a 30~kB message when it has no message in flight, that is, sent but not yet delivered, and when at least six steps, or 0.6~s, have passed since its last send.
The 60~m arena with the radio settings of the 5G-LENA scenario of Section~\ref{sec:eval-lena} keeps every robot inside the SNR range that the 5G-LENA runs covered, and all models share every random draw of the task.
We repeat the run in two heavier settings, namely one message per 0.2~s at $8 \times 16$, called the 0.2~s setting, and 32 robots per environment at 0.6~s, called the 32-robot setting.
In each setting, the i.i.d. delay model draws independent log-normal delays fitted to the delays and losses of the NR engine on the same seeds, and an ideal link delivers every message at once.
A second variant, the empirical i.i.d. model (L0-emp in the tables), draws each delay independently from the pooled delays of the NR engine on the same seeds, so it keeps the whole tail of that distribution, while its loss is fitted as before.
Two models that share seeds also share robot trajectories, so their pooled delay distributions can look closer than two independent samples of one model.
Under load, we therefore compare the even seeds of each model with the odd seeds of ns-3.
The seed-to-seed noise level, given in the Seed noise rows of Table~\ref{tab:closed}, is the KS distance between the even and odd seeds of the NR engine and of ns-3, which reflects sampling noise alone.
Over the five seeds, the 95\% intervals of the AoI p95 reach $\pm$20\% and $\pm$21\% for the NR engine and ns-3 in the 0.2~s setting and $\pm$10\% and $\pm$15\% in the 32-robot setting, and stay within $\pm$13\% elsewhere.

\begin{figure}[t]
\centering
\includegraphics[width=\columnwidth]{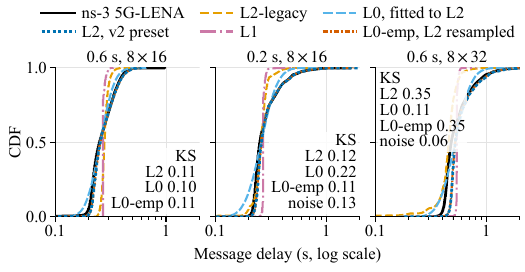}
\caption{Per-message delay CDFs of one scripted controller over each network model in the three traffic settings of Table~\ref{tab:closed}, with the KS distances of L2, L0, and L0-emp to ns-3 in each panel.}
\label{fig:closed-cdf}
\end{figure}

\myparatight{At light load, the network does not shape the traffic}
At light load, nearly every message arrives before the robot's next send, so every model sends almost all of the 6,400 messages that the send rule allows.
We observe in Table~\ref{tab:closed} that the NR engine with the v2 preset and ns-3 agree in delay and AoI, with a pooled KS distance of 0.11.
Figure~\ref{fig:closed-cdf} shows that the fluid model and the prototype engine squeeze almost all delays into a narrow band, with KS distances of 0.58 and 0.46 to ns-3, respectively.
In each panel, the curve of the empirical i.i.d. model lies on that of the NR engine, which it resamples, and the ideal link is omitted because it delivers every message at 0~ms.
The i.i.d. delay model lies at a KS distance of 0.10 to ns-3, and the empirical i.i.d. model lies at 0.11 with the AoI p95 of ns-3, so at light load an independent draw per message is as close to ns-3 as the engine.

\myparatight{Under load, a log-normal delay misses what the controller sees}
In both heavier settings, a robot waits 0.25--1~s for its previous message before it sends the next one.
The robots therefore send 11,000--11,400 messages over the NR engine and ns-3, against 19,200 and 12,800 over the ideal link, so the offered traffic now depends on the network model.
In the 0.2~s setting, the NR engine lies at a KS distance of 0.12 from ns-3, at the seed-to-seed noise level of 0.12--0.13, while the i.i.d. delay model lies at 0.22.
The difference sits in the tail, where the i.i.d. delay model reaches a p95 delay of 404~ms against 500~ms for ns-3.
The controller sees this tail, as its 95th-percentile AoI under the i.i.d. delay model is 29\% lower than under ns-3, outside the seed interval of ns-3, and it sends 9.6\% more messages, since a short delay frees the robot to send sooner.
In the 32-robot setting, neither model reproduces the delay distribution of ns-3.
The median delay of the NR engine lies 9\% above that of ns-3, at 525 against 483~ms, which gives a KS distance of 0.35 against a seed-to-seed noise level of about 0.05, for a reason not yet isolated.
The i.i.d. delay model has the lower KS distance of 0.11 because its wide log-normal CDF cuts through the middle of the ns-3 CDF, even though it misses the tail, with a p95 of 811 against 979~ms.
On what the controller observes, the NR engine matches ns-3 within the seed intervals, with a 95th-percentile AoI of 2.00 against 1.94~s and 11,235 against 11,353 messages, whereas the i.i.d. delay model gives a 33\% lower AoI p95 and 4.5\% more messages.
The prototype engine has a KS distance of 0.28 here, yet its AoI p95 of 1.06~s and its 12,720 messages miss ns-3 by 45\% and 12\%, so its narrow delays do not help the controller.

\myparatight{Independent draws from the engine's own delays still miss the AoI tail}
A log-normal distribution fitted to the median and spread of the NR engine's delays cannot reach their tail, so part of the miss of the i.i.d. delay model may come from its shape, and the empirical i.i.d. model separates this cause from the independence of the draws.
We observe that the empirical i.i.d. model lies as close to ns-3 in delay as the NR engine does, with KS distances on separate seeds of 0.11 and 0.35, and sends within 0.2\% and 0.6\% of the messages sent over ns-3 in the two settings.
However, its 95th-percentile AoI is still 18\% and 16.5\% lower than under ns-3, against 29\% and 33\% lower for the i.i.d. delay model and $-2.0$\% and $+3.1$\% for the NR engine.
Thus, the log-normal shape caused all of the excess messages of the i.i.d. delay model and about half of its AoI miss, and independence causes the rest.
We also count the robot-steps whose AoI exceeds the AoI p95 of ns-3, rounded up to the 0.1~s step of the AoI, which is why fewer than 5\% of the robot-steps of ns-3 exceed it.
This share is 4.2--4.4\% under the NR engine and ns-3 but only 1.3--1.4\% under the empirical i.i.d. model, so its AoI tail is about three times too light in each of the five seeds in both settings.
On the AoI p95 alone, the 18\% miss lies inside the seed interval of ns-3, and the 16.5\% miss lies just outside it.
This is because, under the NR engine and ns-3, 64--70\% of the delay variance lies between robots, and the successive delays of one robot stay close to that robot's own mean, with a correlation of 0.86--0.94 between consecutive messages.
A robot that draws long delays therefore keeps drawing them, and its AoI grows over several messages, whereas independent draws spread the long delays over all robots.
Moreover, the closed-loop delay distribution is known only after a per-slot model has run under the same controller, since the offered traffic depends on the network model.

\myparatight{The task return does not separate the models}
The task return, that is, the summed reward of an episode, is 81.8--82.9 for every model in every setting, including the ideal link, and the robots spend at most 0.26\% of their steps inside a hazard.
Thus, the return of this controller does not depend on the network model in any tested setting, and the comparison rests on delay, AoI, and messages sent.
At light load, the NR engine on the graph backend requires 42~ms per step against 19~ms for eight ns-3 processes, so its speed advantage appears only at the larger scales of Section~\ref{sec:eval-isaac}.
These results underscore that, once the send rule reacts to queueing, the engine reproduces the AoI tail that the controller observes, while an independent delay per message misses it even when drawn from the engine's own delays.

\section{Related Work}
\label{sec:related}

\myparatight{Robot and network co-simulation}
To date, most studies that combine robots with realistic networks couple two simulators through middleware.
For instance, ROS-NetSim~\cite{calvofullana2021rosnetsim} synchronizes a robot simulator with a network simulator through ROS.
CORNET~2.0~\cite{acharya2022cornet2}, SynchroSim~\cite{dey2022synchrosim}, and GazeboNS3~\cite{cauwe2025gazebons3} co-simulate Gazebo robots or drone swarms with ns-3 or emulated Wi-Fi.
CAVIAR~\cite{borges2024caviar} and SimART~\cite{yan2026simart} add ray tracing and 3D scenes for 6G studies.
Closest to our setting, IsaacSim5G~\cite{raghavendra2026isaacsim5g}, as presented in a recent workshop talk, couples Isaac Sim with the OAI 5G stack and ROS~2 for one robot.
It runs at a fixed real-time factor, that is, a fixed ratio of simulated time to wall-clock time, and lists multi-robot orchestration as future work.
Other works pair robots with a network digital twin~\cite{sanchez2025ndt5grobots,lyu2026lucid} or a wireless world model~\cite{madushanka2026coupled}.
Open Wireless Digital Twin~\cite{iye2025openwdt} couples the OAI 5G stack with Sionna ray tracing to emulate one mobile UE in real time, but it includes no robot learning.
The authors in~\cite{solc2026mecamr} train the perception and control models of a mobile robot in Isaac Sim and deploy them over a private 5G network with edge computing, yet the network does not enter the training loop.
Mature cellular simulators such as Simu5G~\cite{nardini2020simu5g} and the Vienna 5G simulators~\cite{muller2018vienna5g}, and vehicular couplings such as Veins~\cite{sommer2011veins}, are accurate, yet none advances thousands of independent scenarios in lockstep with a physics engine under closed-loop traffic and mobility.
However, each of these systems either keeps the network out of training or runs one robot scene against one network instance at or below wall-clock speed.
In contrast, \alg trades a real 5G stack for a per-slot model that runs thousands of environments on one GPU and keeps the network in the loop for about one million robots.

\myparatight{GPU and fast network simulation}
Recently, GeDES~\cite{li2026gedes} runs discrete-event simulation of large wired networks on the GPU, and DONS~\cite{gao2023dons} parallelizes it on CPUs with a data-oriented design and reproduces the event timestamps of ns-3.
On the wireless side, Sionna~\cite{sionna2026sys,hoydis2023sionnart} provides GPU ray tracing and a batched system-level abstraction with link adaptation and PF scheduling.
NVIDIA Aerial Omniverse Digital Twin (AODT)~\cite{nvidia2026aodt} simulates city-scale 5G deployments with GPU ray tracing and GPU-accelerated Radio Access Network (RAN) layers.
However, GeDES and DONS target one large wired network, AODT targets one detailed deployment, and packet-level 5G simulators such as 5G-LENA~\cite{patriciello2019lena} run one scenario per CPU process.
\alg complements these tools, since it reuses the BLER tables of Sionna, validates against 5G-LENA and OAI, and adds the per-robot queues, HARQ timing, and lockstep coupling that a learning loop requires.

\myparatight{Multi-robot learning with communication}
GPU-parallel simulators such as Isaac Gym~\cite{makoviychuk2021isaacgym}, Isaac Lab~\cite{mittal2025isaaclab}, ManiSkill3~\cite{tao2024maniskill3}, and MuJoCo Playground~\cite{zakka2025playground} train policies in thousands of environments at once.
VMAS~\cite{bettini2022vmas} and WarpDrive~\cite{lan2021warpdrive} vectorize multi-agent simulation and learning on the GPU.
DIAL~\cite{foerster2016dial}, CommNet~\cite{sukhbaatar2016commnet}, and TarMAC~\cite{das2019tarmac} learn what to send over channels that deliver every message, and IC3Net~\cite{singh2019ic3net} and SchedNet~\cite{kim2019schednet} learn when or which agents may send.
Recent frameworks inject delay, loss, and bandwidth limits~\cite{purves2026calf,bansal2026agentcommbench} or run several ns-3 environments on the CPU~\cite{burmeister2025defiance}.
QNet~\cite{agarwal2026qnet} learns when to send over a shared network whose state reflects the traffic of other agents.
However, the GPU simulators treat communication as ideal or as an independent delay, and most learning frameworks draw impairments that do not depend on what other agents transmit.
QNet models the shared network, yet without per-slot scheduling or HARQ.
Our \alg instead lets delay and loss emerge from scheduling and retransmission on a shared uplink in every parallel environment, without changing the physics engine.

\section{Conclusion}
\label{sec:conclusion}

In this paper, we introduced \alg, a GPU-batched 5G NR module that brings network-in-the-loop training to massively parallel robot learning, advancing thousands of environments slot by slot in lockstep with Isaac Lab physics.
Fixed-shape network state enables per-environment resets and fast backends verified step by step against a reference backend.
Extensive experiments confirm that the configurable NR engine, once it models the scheduler and the buffer-report grant procedure of ns-3 5G-LENA, reproduces its median delay from light load to saturation with no fitted parameter.
In closed loop, the engine also reproduces the AoI that a controller observes over ns-3, whereas an independent delay per message leaves the AoI tail about three times too light.
In the scale configuration, which is also validated against 5G-LENA, 1,048,576 robots run inside Isaac Lab under a random policy at 1.24 control steps per second on one GPU, against 1.50 without the network.
\alg, its test suite, the co-simulation bridges, and the fleet task are available at \url{https://github.com/ZzZTripleZzZ/isaac-net}, with documentation at \url{https://isaacnet.zifanzhang.com}.

\bibliographystyle{IEEEtran}
\bibliography{ref}

\end{document}